\documentclass[iop]{emulateapj}

\shorttitle{Galaxy environment of a QSO at $z \sim 5.7$}
\shortauthors{Ba\~{n}ados et al.}
\slugcomment{Published in the Astrophysical Journal}

\begin{document}

\title{THE GALAXY ENVIRONMENT OF A QSO AT $z \sim 5.7$}

\author{
Eduardo Ba\~{n}ados \altaffilmark{1},
Bram Venemans \altaffilmark{1},
Fabian Walter \altaffilmark{1},
Jaron Kurk \altaffilmark{2},
Roderik Overzier \altaffilmark{3,4},
}

\author{
Masami Ouchi \altaffilmark{5,6}}

\def\degg{\hbox{$\null^\circ$\hskip-3pt .}}
\def\asec{\hbox{\arcsec\hskip-3pt .}}

\altaffiltext{1}{Max Planck Institut f\"ur Astronomie, K\"onigstuhl 17, 69117, Heidelberg, Germany; \email{banados@mpia.de}}
\altaffiltext{2}{Max Planck Institut f\"ur Extraterrestrische Physik,  Giessenbachstrasse, 68165, Garching, Germany}
\altaffiltext{3}{Department of Astronomy, The University of Texas at Austin, 2515 Speedway, Stop C1400, Austin, TX 78712-1205, USA }
\altaffiltext{4}{Observat\'orio Nacional, Rua Jos\'e Cristino, 77. CEP 20921-400, S\~ao Crist\'ov\~ao, Rio de Janeiro-RJ, Brazil }
\altaffiltext{5}{Institute for Cosmic Ray Research, University of Tokyo, Kashiwa 277-8582, Japan }
\altaffiltext{6}{Kavli Institute for the Physics and Mathematics of the Universe (WPI), The University of Tokyo, 5-1-5 Kashiwanoha, Kashiwa, Chiba 277-8583,
Japan}

\begin{abstract}
High-redshift quasars are believed to reside in massive halos in the early universe and should therefore be located in fields
with overdensities of galaxies, which are thought to evolve into galaxy clusters seen in the local universe. However, despite many efforts, the relationship between galaxy overdensities and
$z\sim6$ quasars is ambiguous. This can possibly be attributed to the difficulty of finding galaxies with accurate redshifts in the vicinity of $z\sim6$ 
quasars. So far, overdensity searches around $z\sim6$ quasars have been based on studies of Lyman break galaxies (LBGs), which probe a redshift 
range of $\Delta z\approx1$. This range is large enough to select galaxies that may not be physically related to the quasar. We use deep narrow- 
and broadband imaging to study the environment of the $z=5.72$ quasar \mbox{ULAS J0203+0012}.
The redshift range probed by our narrow-band selection of Lyman alpha emitters (LAEs) is $\Delta z\approx0.1$, which is significantly narrower than the 
LBG searches. This is the first time that LAEs were searched for near a $z\sim6$ quasar, in an effort to provide clues about the environments
of quasars at the end of the epoch of reionization.
We find no enhancement of LAEs in the surroundings of ULAS J0203+0012 in comparison with blank fields. We explore different explanations 
and interpretations for this non-detection of a galaxy overdensity, including that (1) the strong ionization from the quasar may prevent
galaxy formation in its immediate vicinity and (2) high-redshift quasars may not reside in the center of the most massive dark matter halos.

\end{abstract}


\section{Introduction}

Observations of the highest redshift ($z\gtrsim 6$) quasars indicate that they contain supermassive
black holes with masses $> 10^9$ $M_\odot$ \citep[e.g.,][]{jia07, kur07, der11}. Detecting of such massive black holes less
than a gigayear after the big bang is challenging for structure-formation models. 
Some models propose that supermassive black holes are hosted in high-density peak dark matter halos
\citep[e.g.,][]{vol06}. Numerical models predict  that the most massive dark
matter halos at $z\sim 6$ will evolve into massive $> 10^{14}-10^{15}$ $M_\odot$ 
clusters in the local Universe \citep[e.g.,][]{spr05}.

At redshifts of $2<z<5$, significant galaxy overdensities or protoclusters have been found around more than a half dozen luminous radio galaxies, confirming
the idea that luminous active galactic nuclei (AGNs) pinpoint dense regions in the early universe \citep[see, e.g.,][and references therein]{ven07a}.
Nevertheless, the environments associated with other classes of AGNs, such as the optically selected quasars, are currently less well constrained.

Recently, there have been a number of studies finding contradictory results regarding the galaxy environment of quasars at $z \sim 2$--$5$. 
\cite{can12}, in a narrow-band imaging survey around a quasar at $z=2.4$, found a much larger number of Lyman alpha emitters (LAEs) than 
in blank-field Ly$\alpha$ surveys. They suggest that this overdensity may be fully explained by quasar fluorescence, which boosts
gas-rich but intrinsically faint LAEs, increasing the number of detectable objects. \cite{fra04} carried out deep narrow-band observations
centered on a $z = 2.168$ quasar. They were expecting to see $6$--$25$ Ly$\alpha$-fluorescent clouds
 and tens of normal LAEs, based on similar surveys. However, they did not detect any Ly$\alpha$ emission. 
In light of this result, \cite{bru12} developed a semi-analytical model to interpret the
observations. They concluded that the intense ultraviolet emission of the quasar may be suppressing the 
star formation in galaxies that are situated close to the quasar. This is consistent with the findings of \cite{kas07} that LAEs are clustered around a quasar at $z=4.87$ but 
avoid it in its immediate vicinity ($\sim 4.5$ comoving Mpc). \cite{swi12}, on the other hand, reported a galaxy overdensity an order of magnitude higher than
that which might be expected in the field within $8.2$ comoving Mpc of a quasar at $z=4.528$. The same authors did not find clear evidence of overdensities
in the fields of two quasars at $z\sim 2.2$. Very recently, \cite{hus13} noted
that the luminous quasars in their sample at $z\sim 5$ were typically found in 
overdense regions. Nevertheless, they mentioned that even the richest quasar 
environment they studied was no richer than others structures identified in 
blank fields.

Since the discovery of $z \sim 6$ quasars, several groups have tried to identify galaxy overdensities associated with them.
The results were as puzzling as those at lower redshifts. No unam\-bi\-guous relation has been found between galaxy 
overdensities and $z \sim 6 $ quasars, mainly due to the  difficulty of fin\-ding galaxies with accurate 
redshifts at \mbox{$z \sim 6$.}  So far, these efforts were mainly based on studies of con\-ti\-nuum $i$-dropout galaxies, characterized
by a large magnitude difference between the $i$ and $z$ bands (i.e., using the same technique that was used to detect the quasars).
This technique probes a redshift range of approximately $\Delta z \approx 1$, which is large enough to identify galaxies that are
not physically related to the quasar. \cite{wil05} carried out a survey with Gemini of three $z > 6.2$
quasars. They found no evidence for an overdensity of $i$-dropouts in the $27$ arcmin$^2$ field surrounding the quasars.
\cite{sti05}, however, came to a different conclusion for one of these fields. They observed a small field around SDSS J$1030+0524$
at $z = 6.28$ with the Advanced Camera for Surveys (ACS) of the \textit{Hubble Space Telescope} (\textit{HST}) and found more than twice
the number of dropouts with $i_{775} - z_{850} > 1.5$ expected
from statistics obtained by GOODS. The ACS field used in this study only surveys the region close
to the quasar, but the observations are more sensitive to faint galaxies. \cite{sti05} explain that the difference is that their data set
is deeper and the majority of the excess sources are fainter than the limiting magnitude of \cite{wil05}.
\cite{zhe06}, using ACS observations, found an overdensity in the field of a radio-loud quasar at $z=5.8$: the surface density of $1.3 < i - z < 2.0$
sources was about six times higher than the number expected from ACS deep fields. 
\cite{kim09} studied $i$-dropout galaxies in five fields centered on $z \sim 6$ quasars using ACS and reported that regions near 
quasars are sometimes overdense and sometimes underdense. From a theoretical perspective,
\cite{ove09}, using semi-analytic galaxy models in combination 
with the dark matter Millennium simulation \citep{spr05, del07}, showed that the lack of neighboring galaxies
as found by \cite{wil05} and \cite{kim09} was not inconsistent with quasars occupying massive halos due to a combination of depth,
field-of-view, and projection effects. However, the same simulations also predicted that there
were regions much more overdense than seen around any $z \sim 6$ quasar observed to date.
\cite{uts10} found an overdensity of $i$-dropout galaxies around a quasar at $z=6.43$, but at the same time these
objects avoided the center near the quasar ($\sim 15$ comoving Mpc), similar to what \cite{kas07} found at lower redshift.
At high redshift, it is very hard to obtain spectra of faint galaxies, but there are a
few cases at $z\sim 5$, based on available spectroscopy, where quasars are located in 
regions with an overdensity of galaxies \citep[][]{cap11, wal12, hus13}.
It is worth noting that in some of these cases, the quasars do not reside in 
the center of the overdensities but at distances of $\sim 5$--$15$ comoving Mpc.
It is important to note that a few galaxy overdensities or protoclusters have been
discovered serendipitously in random fields (e.g., Ouchi et al. 2005 at $z \sim 5.7$; Toshikawa et al. 2012 at $z \sim 6$;
Trenti et al. 2012 at $z \sim$ 8), suggesting that not all overdensities host AGNs, although this could also be explained
by the duty cycle of AGN activity.

In summary, current dropout studies at $z \sim 6$ give partly contradictory results. An efficient alternative is to search for LAEs in a 
narrow redshift range near the quasar using narrow-band filters.
However, mainly due to the extremely low density of $z\sim 6$ quasars of $\sim 6 \times 10^{-10}$ Mpc$^{-3}$ \citep{fan04}, until recently
no high-redshift quasars were known to have a redshift that shifts the Ly$\alpha$ line into a region of the optical 
spectrum that is devoid of bright sky emission lines. These atmospheric windows, for example, around \mbox{$8160$ \AA} $\,$ and $9120$ \AA, have 
successfully been used by blank field Ly$\alpha$ surveys to search for galaxies at $z=5.7$ and $z=6.6$  \citep[e.g.,][]{hu99,hu02,ouc05,ouc08}.
So far, the only study using narrow-band imaging to detect the Ly$\alpha$-emitting
halo and possibly companions around quasars at $z>6$ was carried out using the
\textit{HST} by \cite{dec12}. Even though the goal of these observations was to detect
the Ly$\alpha$ halo around the quasars, they did not find any companions in the 
immediate vicinity of two $z>6$ quasars.  This latter result was not unexpected 
because of the small field-of-view covered by their study  ($\sim 1$ arcmin$^2$).

In the present work, we use deep narrow- and broad-band imaging to study the 
environment of the broad-absorption line quasar  ULAS J0203+0012
\citep[hereafter J0203;][]{ven07b, jia08,mor09}, one of the first $z \sim 6$ 
quasars known with a redshift that shifts the Ly$\alpha$ line into an
atmospheric window that allows for deep narrow-band imaging.
We assume the quasar redshift of  $z = 5.72$ determined by \cite{mor09}, which is based
on the N$\,${\sc v}, Si$\,${\sc iv}, C$\,${\sc iv}, and C$\,${\sc iii} lines
from a combined optical and near-infrared spectrum. However, we have to keep in mind that any redshift in the range of $5.70 < z < 5.74$ is consistent
with their data. \cite{rya09}, using a Keck/NIRSPEC spectrum, independently confirmed this quasar, reporting a redshift of $z=5.706$ based on
a broad emission feature that they presume to be C$\,${\sc iv} $\lambda\, 1549.062$. In any case, the uncertainty in redshift does not
shift the Ly$\alpha$ line outside the narrow-band filter used in our work.

In Section \ref{ss:obs} we describe the data used in this study, photometry, and object detection.
Section \ref{sec:selection} describes how LAE and dropout or Lyman break galaxy
(LBG) candidates were selected. In Section \ref{sec:results}
we present our results, including a number count comparison with blank fields,
photometric properties of the LAE sample, and our estimation of the black hole mass
of the quasar. Finally, in Section \ref{sec:discussion} we discuss the results
and present our conclusions.

In this paper, all magnitudes are given in the AB system and are corrected for Galactic extinction \citep{sch11}.
We employ a $\Lambda$CDM cosmology
with $H_0 = 70 \,\mbox{km s}^{-1}$ Mpc$^{-1}$, $\Omega_M = 0.3$, and $\Omega_\Lambda = 0.7$, which
yields an age of the universe of $0.976$ Gyr and a spatial scale of $39.9$ kpc arcsec$^{-1}$
in comoving units at $z=5.7$.

\section{Data, Reduction and Photometry} \label{ss:obs}

\begin{figure}
\epsscale{1.0}
\plotone{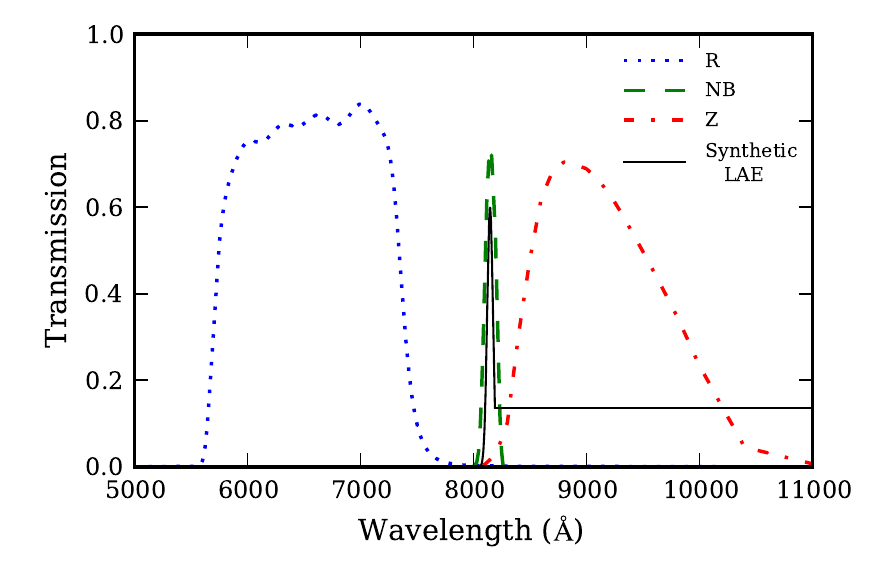}
\caption{Transmission curves of the filters used in this work and
the synthetic spectrum of a LAE at $z=5.7$.\label{fig:filters}}
\end{figure}

The field centered on the quasar J0203 at $z=5.72$ was observed during 2010  November--December.
Narrow- and broad-band imaging was carried out with the FOcal Reducer/low dispersion Spectrograph 2 \citep[FORS2;][]{app92} using the 
red sensitive detector consisting of two 2k $\times$ 4k MIT CCDs  at the Very Large Telescope (VLT). 
The pixels were $2 \times 2$ binned to decrease the readout time and noise, giving a spatial scale of $0.25$ arcsec pixel$^{-1}$. 
The field-of-view was $6\arcmin.8 \times 6\arcmin.8$. The field was observed in  the 
narrow-band filter (FILT\_815\_13, $\lambda_c = 8150$ \AA, $\Delta \lambda = 130$ \AA, hereafter 
NB) and the broad-band filters
$Z$ (Z\_GUNN, $\lambda_c = 9100$ \AA, $\Delta \lambda = 1305$ \AA) and $R$ (R\_SPECIAL, $\lambda_c = 6550$ \AA, $\Delta \lambda = 1650$ \AA).
The filter transmission curves are shown in Figure \ref{fig:filters}. This filter set allowed us to sample
both LAEs in a redshift range of $5.66 \lesssim z \lesssim 5.75$ with the narrow-band filter and LBGs in a broader redshift range
$5.2 \lesssim z \lesssim 6.8$ with the broad-bands ($5.2 \lesssim z \lesssim 5.8$,  if we add the narrow-band filter to the selection).

The individual exposure times in NB, $Z$, and $R$ were 800s, 110s, and 240s per pointing, respectively. 
The individual exposures were shifted by $\sim 10 \arcsec$ with respect to each other to facilitate the removal of
bad pixels and flat field errors associated with a fixed position of the CCD pixel. The total integration times 
were 6 hr in NB, 1.5 hr in $Z$, and 48 minutes in $R$. 

Standard data reduction was performed, which consisted of bias subtraction, flat fielding,
sky subtraction, image alignments, and stacking.
We calculated the photometric zero points by using the magnitudes and colors of stellar objects
in the Sloan Digital Sky Survey (SDSS) catalog \citep{aba09}. With this procedure,
the flux losses for point sources are corrected, which is appropriate for this study
since we are interested in high-redshift LAEs that are expected to be unresolved.
The accuracy of the photometric zero points was $0.04$, $0.04$, and $0.05$ for the NB, $Z$, and $R$ frames, respectively. 

The area covered by the reduced images was $\sim 44.4$ arcmin$^2$. 
The seeing for the final NB, $Z$, and $R$ images was $0\asec84$, $0\asec86$, and $0\asec76$ respectively.
The $5\sigma$ limiting magnitudes of the reduced images with a $1\asec5$ 
diameter aperture were $\mbox{NB} = 25.34$,  
$Z = 25.14$, and $R = 26.29$.

In order to determine appropriate colors, the NB and $R$ images were convolved to match the PSF of the $Z$ image, the image with
the worst seeing, using the IRAF task \textit{psfmatch}.  The NB total magnitudes were calculated from the unconvolved image.

The source catalog was created running SExtractor \citep{ber96} in dual image mode with the 
narrow-band frame as the detection image.
Aperture magnitudes were calculated using a $1\asec5$ diameter aperture 
($\sim1.75 \times$FWHM of the seeing). The chosen aperture was a good balance
between optimizing the signal-to-noise (S/N) photometry of point sources and minimizing the amount of flux outside the 
aperture \citep{lab03}. This size assured that at least $70\%$ of the flux of a point
source was inside the aperture.  We also tested larger apertures (e.g., 1\asec8 diameter)
and the results did not change, but the S/N of the objects slightly decreased. 
Magnitudes of objects not detected or fainter than $2\sigma$ limiting magnitudes either
in $Z$ or $R$ were replaced by the corresponding $2\sigma$ limiting magnitude. Finally,
we only considered objects with S/N greater than 4, objects with 
$\mbox{NB} > 18$, and with SExtractor flag $\leq 4$ 
in order to eliminate objects flagged as truncated (too close to an image 
boundary), incomplete or corrupted. The final catalog contained 2424 objects.

In order to conclude whether or not an overdensity around the quasar was present,
we needed a comparison field obtained at similar depth in a region that did not
target a $z \gtrsim 5.7$ quasar. 
For LAEs we were able to use results from the literature (see Section \ref{ss:LAEcomp}), whereas for LBGs there
were no $R-Z$ dropout searches with large spectroscopic follow-up in the literature.
We performed our own selection of $R-Z$ dropouts using the public catalogs of the 
Subaru Deep Field (SDF) imaging survey \citep{kas04}, which (after removing low-quality
regions) had an effective area of $\sim 876$ arcmin$^2$.
The SDF had several characteristics that made it a suitable comparison field for our study.
(1) It is a large field, which helps to reduce cosmic variance. 
(2) It has $R$ and $Z$ bands covering similar wavelength ranges compared to the ones in FORS2. 
(3) There are 42 spectroscopically confirmed $i$-dropout galaxies discovered in different 
studies \citep{nag04,nag05, nag07, ota08, jia11, tos12, jia13}, which can be directly used to 
quantify the accuracy of our LBG selection.

The confirmed $i$-dropout galaxies in the SDF were originally selected in the $Z$-band and 
all of them have a redshift $z > 5.9$ (with one exception at $z=5.762$). For that reason, these galaxies would either not be detected
or have negligible flux in our NB filter. Thus, in order to 
perform a consistent comparison, we created a second catalog for the J0203 field 
this time using the $Z$-band frame as the detection image (see Section \ref{sec:LBGcomp}). 

\section{Candidates Selection} \label{sec:selection}

Figure \ref{fig:cmd} shows the two-color diagram of $Z - \mbox{NB}$ and $R - Z$ for the objects detected in the narrow-band
filter and summarizes the selection criteria for LAE and dropout candidates.

\subsection{LAEs} \label{ss:laesel}
LAEs are a population of high-redshift galaxies whose spectra are dominated by 
a strong Ly$\alpha$ emission line and have a very flat and faint continuum. The LAE candidates were selected according to the following criteria:

\begin{itemize}
 \item \textit{Narrow-band excess.} A positive $Z - \mbox{NB}$ indicates an excess in the narrow-band flux intensity.
We determine a color cut such that the flux in the narrow-band is twice that in the $Z$ band: $Z - \mbox{NB} > 0.75$. This cut corresponds to
selecting objects with a rest-frame equivalent width greater than $17.7$ \AA $\,$(see Section \ref{sec:LAEprop}).
 \item \textit{Continuum break.} To differentiate between high-redshift LAEs and low-redshift emission line
interlopers, we require a break in the continuum: $R - Z > 1.0$.
 \item \textit{Significance of the narrow-band excess.} To avoid contamination by objects that satisfy the color criteria only due
to photometric errors, we require: $\left| (Z - \mbox{NB})\right| > 2.5 \sqrt{\sigma^{2}_{Z} + \sigma^2_{\mbox{{\scriptsize NB}}}} $.
\end{itemize}

As seen in Figure \ref{fig:cmd}, there is one object that satisfies our criteria 
(LAE 1). There is another one that is not detected in either of the broad-band images (LAE 2).
LAE 2 has a lower limit in the narrow-band excess of $Z-\mbox{NB} > 0.62$. 
Additionally, when visually inspecting the $Z$-band image of LAE 2 a faint source is apparent.
Forced photometry in the $Z$-band image at the position of LAE 2 gives a 
$1.9\sigma$ signal, thus, increasing the chances of LAE 2 being a real object. 
We consider LAE 2 to be a tentative LAE candidate, although deeper observations and/or 
spectroscopy are needed to confirm its nature.

\subsection{Lyman Break Galaxies} \label{ss:LBGsel}

Dropout galaxies are also known as LBGs, given that they are selected
using the Lyman break technique \citep{ste96}. The basis of this method is the fact that 
hydrogen is very effective at absorbing radiation at
wavelengths shorter than $912$  \AA $\,$ (the Lyman limit). Therefore, radiation
with $\lambda \lesssim 912$ \AA $\,$ is strongly suppressed by
intergalactic and interstellar absorption, so a very small fraction of these photons
will reach us, forming the so-called Lyman Break.  At high-redshifts, the Ly$\alpha$ forest
becomes so optically thick that a large fraction of the light coming from $\lambda_{\mbox{{\scriptsize rest}}}= 912 $ \AA $\,$ to $\lambda_{\mbox{{\scriptsize rest}}} = 1216 $ \AA $\,$
is also absorbed. At this point the Lyman break selection becomes, in effect, 
a selection of objects with a sharp break at $\lambda_{\mbox{{\scriptsize rest}}} = 1216$  \AA.

Dropout candidates were selected with the following criteria:

\begin{itemize}
 \item \textit{Continuum break.} A prominent break in the continuum is expected for high-redshift galaxies due
to the hydrogen absorption. We require a continuum break of $R - Z > 2.0$. Additionally, 
in an attempt to bias our high-redshift candidates
toward a redshift closer to the redshift of the quasar, we require
a break between the flux in the NB and $R$ filters: $R - \mbox{NB} > 1.5$. However,
this last criterion has a small effect and excludes only one candidate. 
 \item \textit{Significance of the break.} To avoid contamination by objects that only satisfy the color criteria due
to photometric errors, we require: $\left| (R - Z)\right| > 2.5 \sqrt{\sigma^{2}_{R} + \sigma^2_{Z}} $.
 \item \textit{Faintness.} Since high-redshift galaxies are expected to be faint, we 
 impose a minimum magnitude of $Z>21$.
\end{itemize}

\begin{figure}
\epsscale{1.0}
\plotone{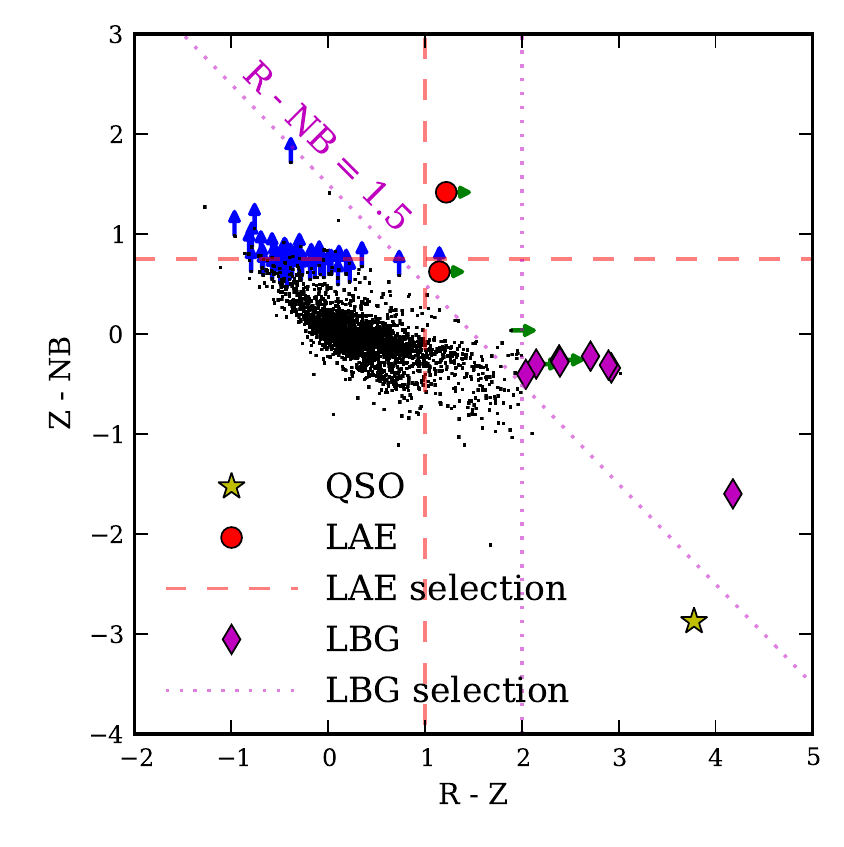}
\caption{Two-color diagram with objects detected in the narrow-band filter.
Vertical blue and horizontal green arrows show $2\sigma$ limiting magnitudes in $Z$ and $R$ for objects
not detected in the respective filters. The selection criteria for LAE and dropouts are
explained in Section \ref{sec:selection}.\label{fig:cmd}}
\end{figure}

As shown in Figure \ref{fig:cmd}, there are eight LBG candidates satisfying the criteria.

\begin{figure}
\epsscale{1.0}
\plotone{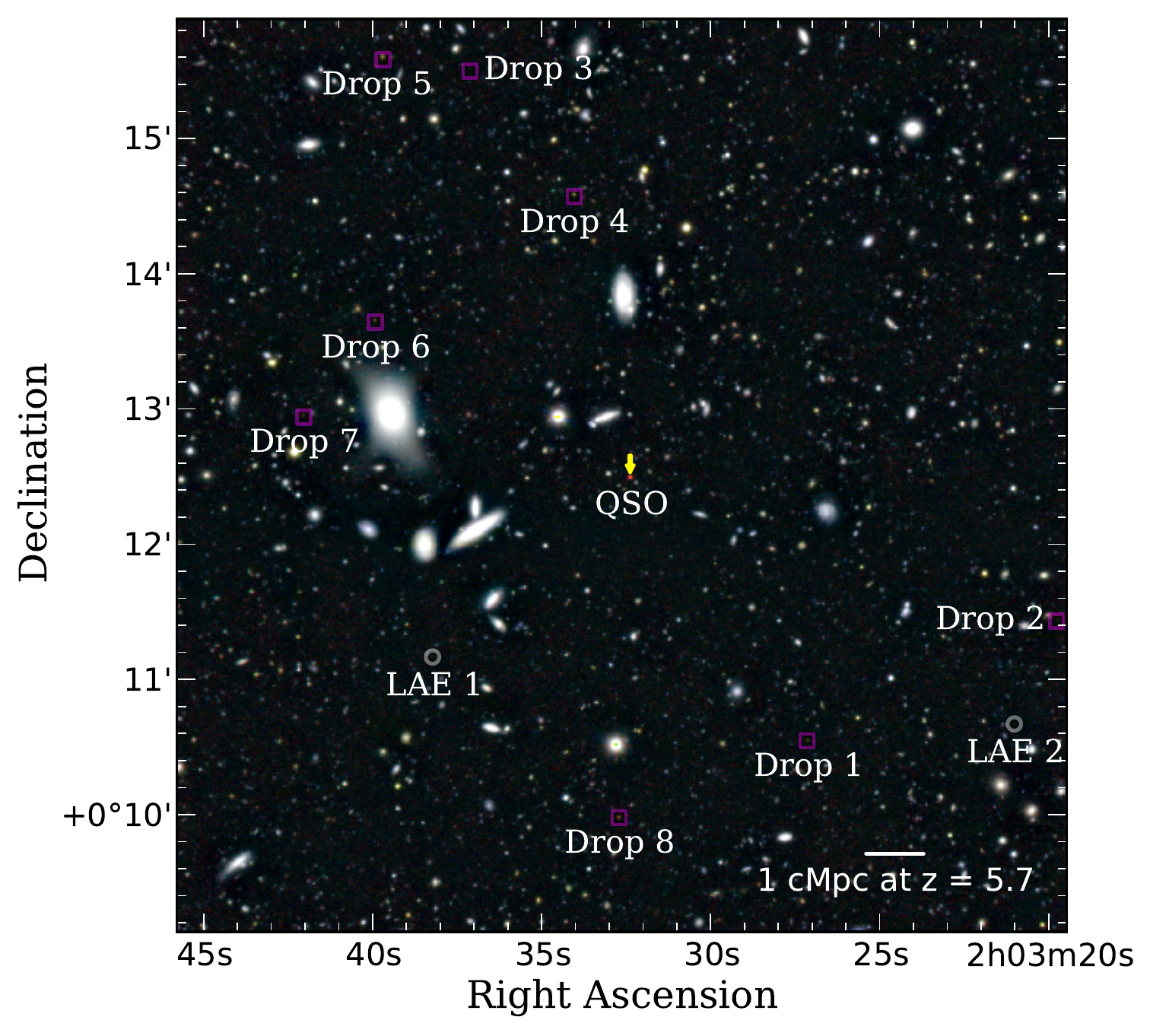}
\caption{Color image of the 44.4 arcmin$^2$ field centered on ULAS J0203+001. 
The objects of interest are highlighted.\label{fig:field}}
\end{figure}

\begin{figure}
\epsscale{0.80}
\plotone{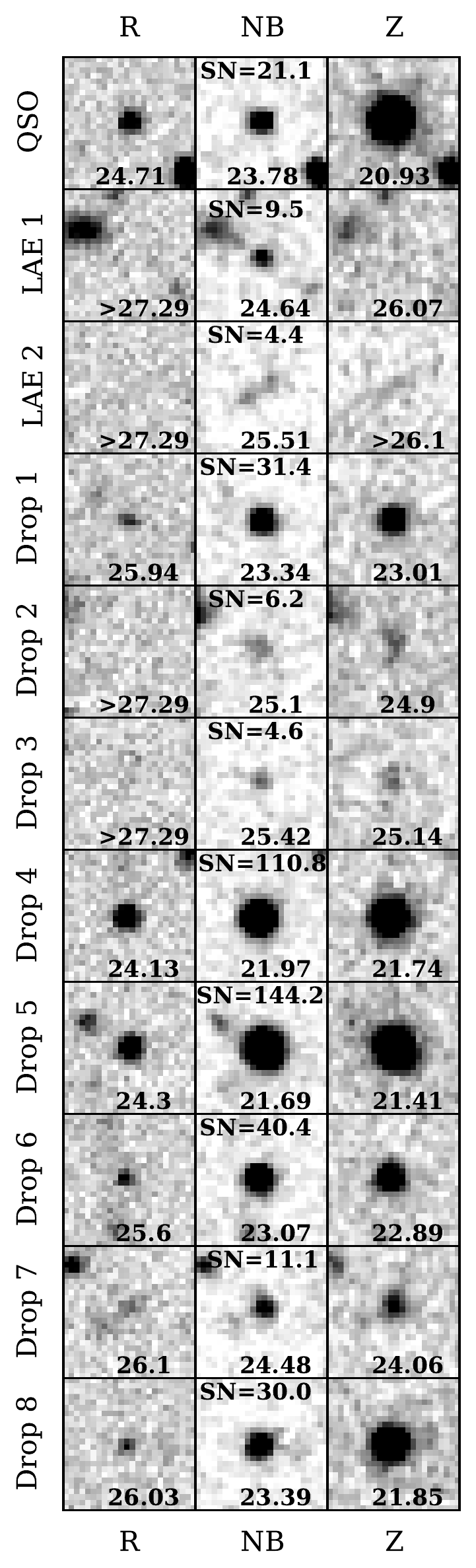}
\caption{Postage stamp images for the QSO and our candidates, each presented in a 
$6^{\prime \prime} \times 6^{\prime \prime}$ box. For every stamp, the respective magnitude 
(or $2\sigma$ limiting magnitude) is given. The signal-to-noise ratio (S/N) in the narrow-band is also shown.\label{fig:stamps}}
\end{figure}

The coordinates of the LAE and LBG candidates and their projected distances to the quasar are presented in Table \ref{tbl-1}.
Figure \ref{fig:field} shows the color image of the field overlaid with the positions of the
LAE and LBG candidates, and the quasar. Postage-stamp images of these objects are shown in \mbox{Figure \ref{fig:stamps}}.

\begin{table}
\begin{center}
\caption{Coordinates of the Candidates of Figure \ref{fig:field} and their Projected Distances
 with Respect to the Quasar\label{tbl-1}}
\begin{tabular}{ccccc}
\tableline\tableline
Object & R.A. & Decl. & Distance & Distance \\ 
       & (J2000.0) & (J2000.0)	 & (arcmin)	& (cMpc$^a$)\\
\tableline
LAE 1  & 02:03:38.23 & +00:11:09.9 & 1.97 & 4.65 \\
LAE 2  & 02:03:21.02 & +00:10:40.2 & 3.37 & 7.96\\
Drop 1 & 02:03:27.15 & +00:10:32.7 & 2.34 & 5.52 \\
Drop 2 & 02:03:19.76 & +00:11:25.8 & 3.33 & 7.85 \\
Drop 3 & 02:03:37.12 & +00:15:29.9 & 3.24 & 7.64 \\
Drop 4 & 02:03:34.04 & +00:14:34.2 & 2.13 & 5.02\\
Drop 5 & 02:03:39.69 & +00:15:35.0 & 3.60 & 8.49\\
Drop 6 & 02:03:39.92 & +00:13:38.5 & 2.21 & 5.22\\
Drop 7 & 02:03:42.04 & +00:12:56.3 & 2.46 & 5.78\\
Drop 8 & 02:03:32.71 & +00:09:58.6 & 2.51 & 5.92\\
\tableline
\end{tabular}
\end{center}
\hspace{1pt} $^a$Comoving Mpc at $z=5.7$

\end{table}

\section{Results} \label{sec:results}

\subsection{Blank Field Comparison} \label{ss:comparison}

\subsubsection{LAEs}\label{ss:LAEcomp}

The selection criteria used in the present work (see Section \ref{ss:laesel}) is close to that used by  \cite{ouc05,ouc08}. Thus, it 
is natural to use their LAE sample for comparison. Their observations cover a larger area and reach fainter luminosities than this work: the area
they imaged is 1.04 deg$^2$ on the sky and a 5$\sigma$ limiting magnitude of $\mbox{NB}=26.0$ ($\lambda_c = 8150$ \AA, $\Delta \lambda = 120$ \AA).
Unlike our field, the  \cite{ouc08} sample is not centered on a quasar, which is
why we consider it a blank field.  However, even in blank fields,
protoclusters can exist. In fact, \cite{ouc05} detected two overdensities that could be
clusters in a formation phase.

\begin{figure}
\epsscale{1.0}
\plotone{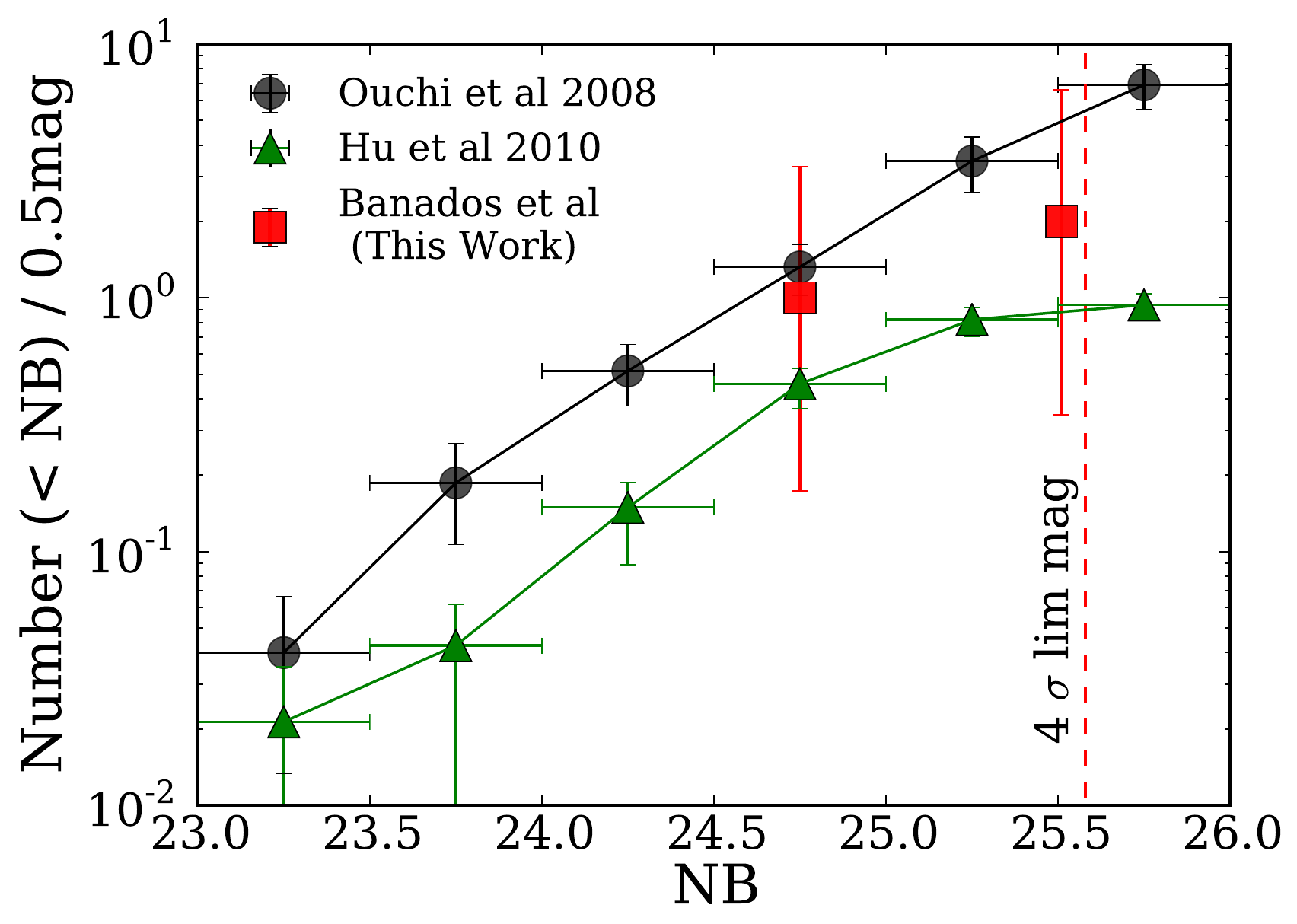}
\caption{Cumulative number of LAEs scaled to our area. Black circles, green triangles, and red squares
correspond respectively to the samples from \cite{ouc08}, \cite{hu10}, and this work. The vertical dashed 
line shows the $4\sigma$ limiting magnitude in the current work.
Our measurements are consistent with no overdensity in our quasar field.\label{fig:blank_fields}}
\end{figure}

\begin{figure}
\epsscale{1.0}
\plotone{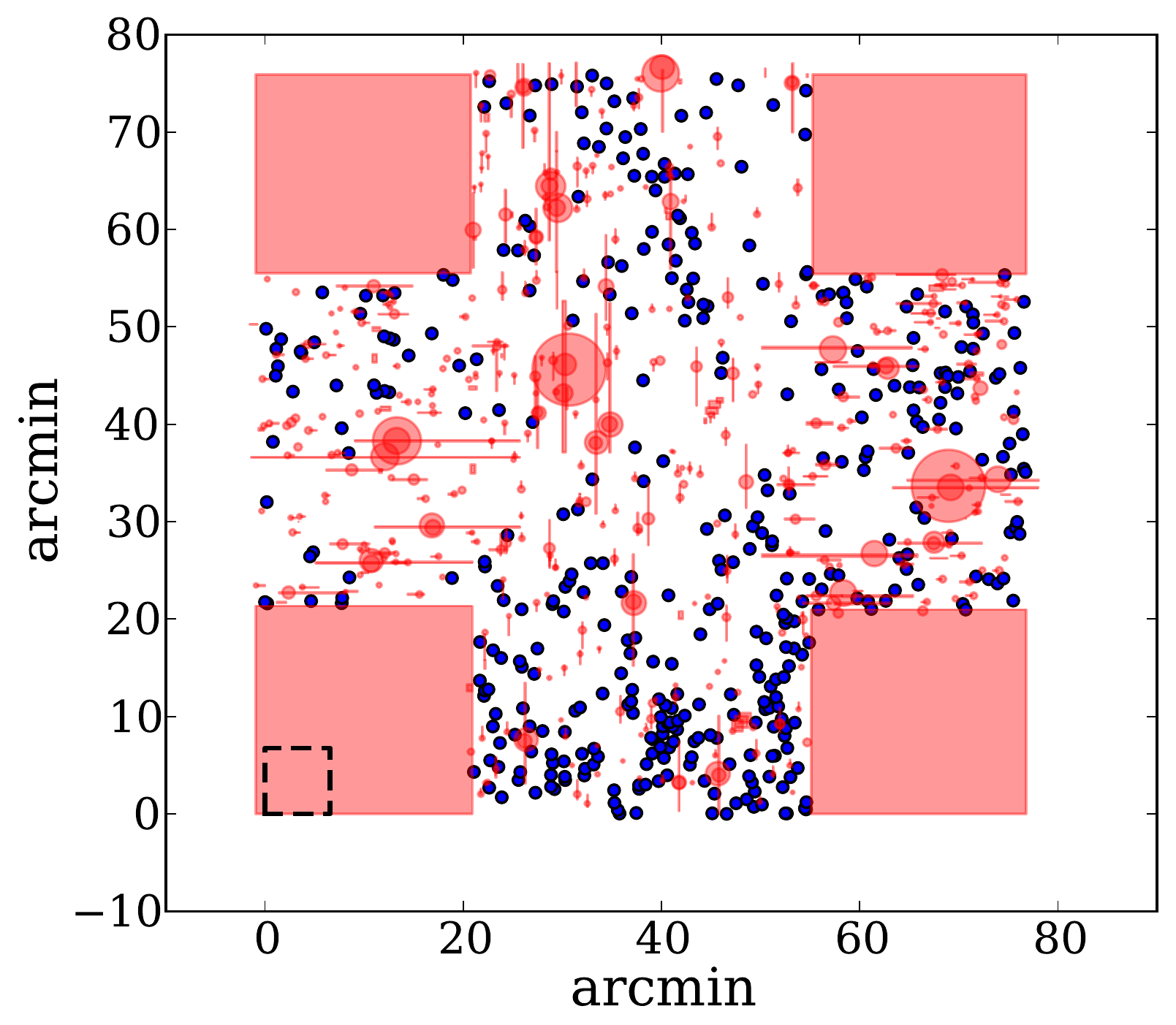}
\caption{LAE candidate distribution found by \cite{ouc08}. Masked regions are shown in red. 
The size of the field-of-view (FORS2) used
in this work is represented by the dashed box in the bottom left corner.\label{fig:ouchi_laes}}
\end{figure}

\begin{figure}
\epsscale{1.0}
\plotone{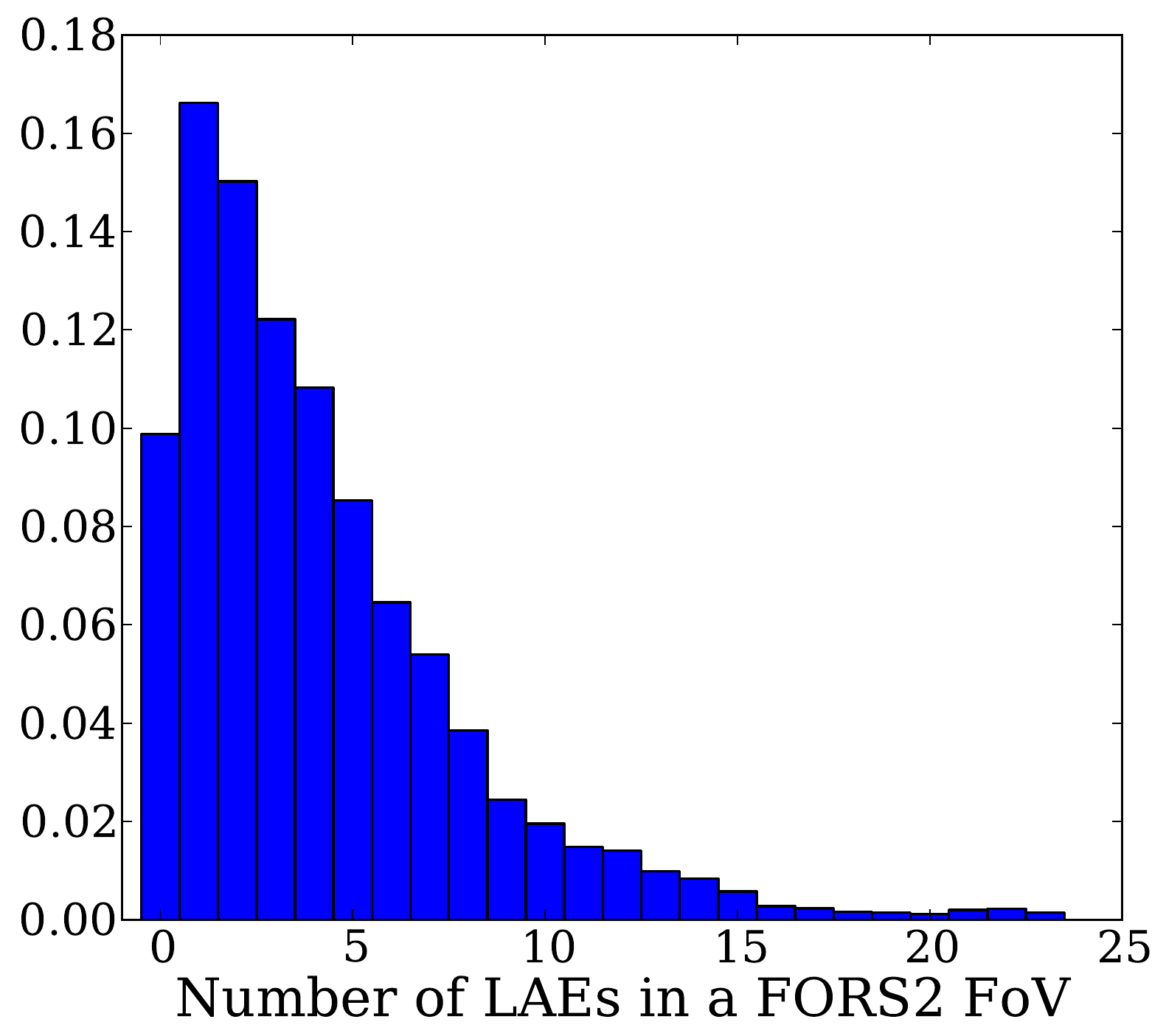}
\caption{Normalized histogram of the number of LAEs found within a FORS2 field-of-view. 
We count the number of LAEs in $100,000$ fields-of-view randomly placed in the \cite{ouc08}
LAE sample (see Figure \ref{fig:ouchi_laes}).\label{fig:ouchi_histo}}
\end{figure}

In Figure \ref{fig:blank_fields}, we show our cumulative number of LAEs and also the numbers from \cite{ouc08} scaled to our area.
Our results are in good agreement with the expected number from \cite{ouc08}.
Figure \ref{fig:ouchi_laes} shows the distribution
on the sky of the 401 LAEs at $z=5.7\pm 0.05$ detected by \cite{ouc08}. Masked regions due to bright stars or image artifacts are shown in red. The dashed square
in the bottom-left corner represents the effective size of the FORS2 field-of-view used in this work. With the goal of estimating the probability of detecting,
in a blank field, the number of LAEs that we find in this work, we placed $100{,}000$ FORS2 fields-of-view at random positions in the \cite{ouc08} field.
Only fields where less than 10\% of the region is masked out were considered. We counted how many LAEs fell in each FORS2 field and the result is summarized in
the histogram of Figure \ref{fig:ouchi_histo}.  Even considering
that \cite{ouc08} would have detected more galaxies since they were sensitive
to fainter magnitudes, we find from Figure \ref{fig:ouchi_histo} that the number
of LAEs in our field-of-view is consistent with the most typical number expected in 
their blank field, i.e., one or two galaxies. 

Our results also compare with the study by \cite{hu10}, which is based on a sample of 88 
spectroscopically confirmed LAEs at $z\sim5.7$ in an area of 1.16 deg$^2$.
Figure \ref{fig:blank_fields} shows their cumulative number of LAEs
scaled to our area. The fact that their numbers are lower than the numbers
of the \cite{ouc08} sample, could be explained by the fact that \cite{hu10} could not
spectroscopically confirm approximately half of their photometric candidates.
Nevertheless, our results are still consistent, within the errors, with the number 
expected from \cite{hu10}. 
We conclude that there is no overdensity of LAEs in the quasar field.

\subsubsection{LBGs}\label{sec:LBGcomp}

As stated in Section \ref{ss:LBGsel} and shown in Figures \ref{fig:field} and 
\ref{fig:stamps}, we found eight objects satisfying our LBG criteria. 
These objects were detected in the NB filter and our criteria 
place them in the redshift range: $5.2 \lesssim z \lesssim 5.8$. 
However, we cannot follow the same approach if we want to exploit the information 
of the 42 confirmed $i$-dropout galaxies in our comparison field (SDF), due to the
fact that essentially all of these confirmed galaxies have a redshift $z>5.9$ 
(except one that has a redshift of $z=5.762$, $\mbox{ID}=1$ in Toshikawa et al. 2012) and would not 
be detected in our NB filter.

We consider the SDF to be a blank field because it does not contain known $z\gtrsim5.7$ quasars.
The only caveat, is that it
has been claimed that the SDF contains two overdensities. One is
at $z\sim 4.9$ \citep{shi03}, from which we expect minimal contamination due to our
selection criteria. The other one is a protocluster at $z\sim 6$ \citep{ota08,tos12},
which in principle could complicate our analysis.
 
The creation of the source catalogs for both the quasar and comparison fields was carried out
in the same manner as for the J0203 field in Section \ref{ss:obs}, but this
time using the $Z$-frame as the detection image.  We applied the same selection criteria 
as in Section \ref{ss:LBGsel}, except for the narrow-band constraints. Additionally,
since the depth of the fields are different, in order to make a consistent comparison,
we have constrained our LBG candidates to have a magnitude brighter than the
$5\sigma$ limiting magnitude in the quasar field, i.e., $Z<25.14$. Since all the 
confirmed members of the protocluster at $z\sim 6$ have magnitudes $Z>25.5$ \citep{tos12}, our 
$Z<25.14$ selection cut should at least prevent us from dealing with a large fraction 
of the protocluster galaxies.

\begin{figure}
\epsscale{1.0}
\centering
\plotone{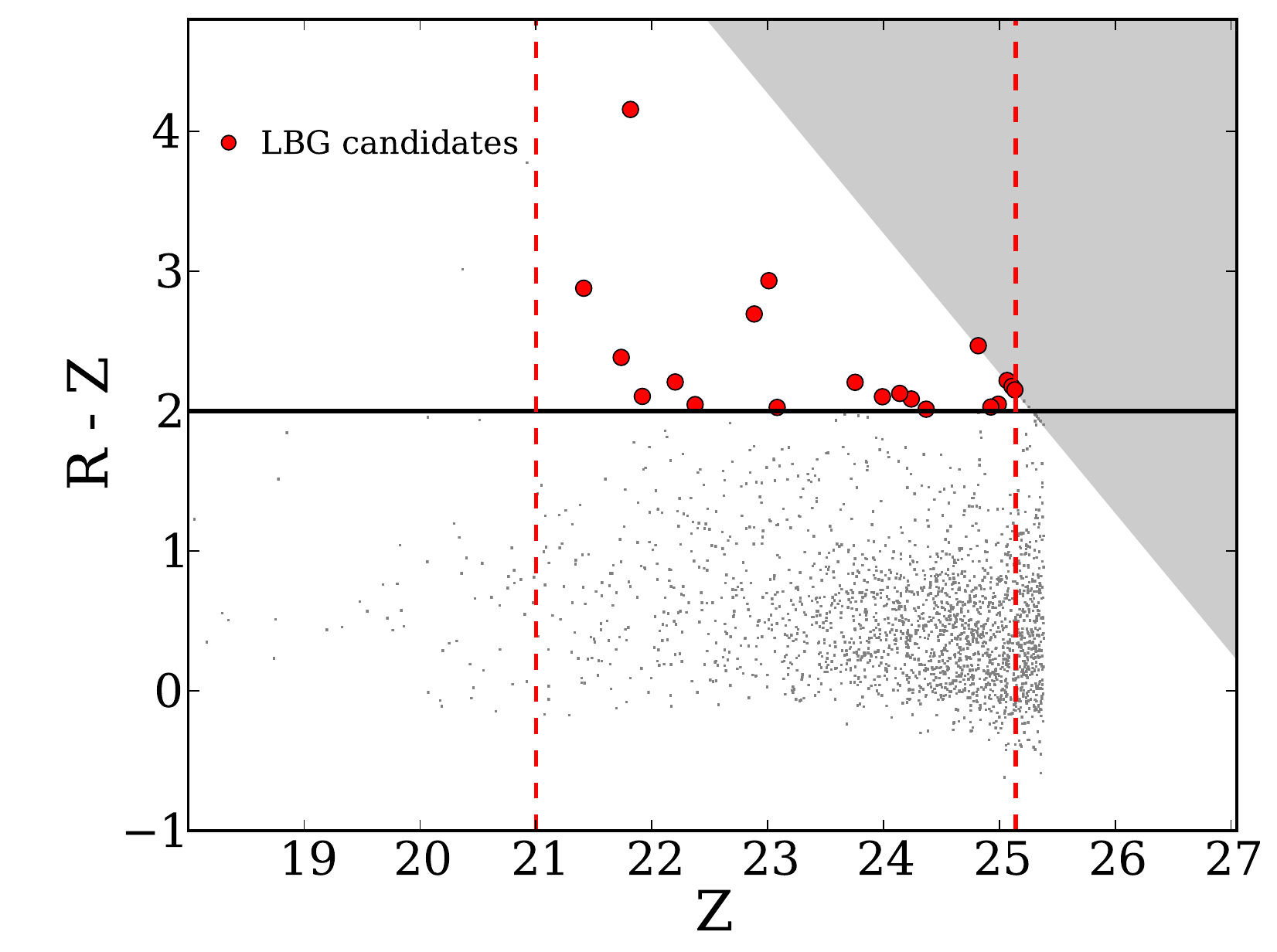}
\caption{Color--magnitude diagram for J0203 field using the $Z$-frame as the detection image. 
Gray points represent all the objects in the field.
The horizontal solid line is our criteria for selecting LBGs with a break larger 
than $R-Z=2$.  The vertical dashed lines are the brighter and fainter magnitudes 
considered for our candidates, $Z= 21$ and $Z=25.14$, respectively (see the text).
The gray region shows $R$-band magnitudes fainter than $2\sigma$ limiting
magnitude ($R=27.29$).  The 20 LBG candidates
are indicated with red circles.\label{fig:cmd_qf}}
\end{figure}

\begin{figure}
\epsscale{1.0}
\centering
\plotone{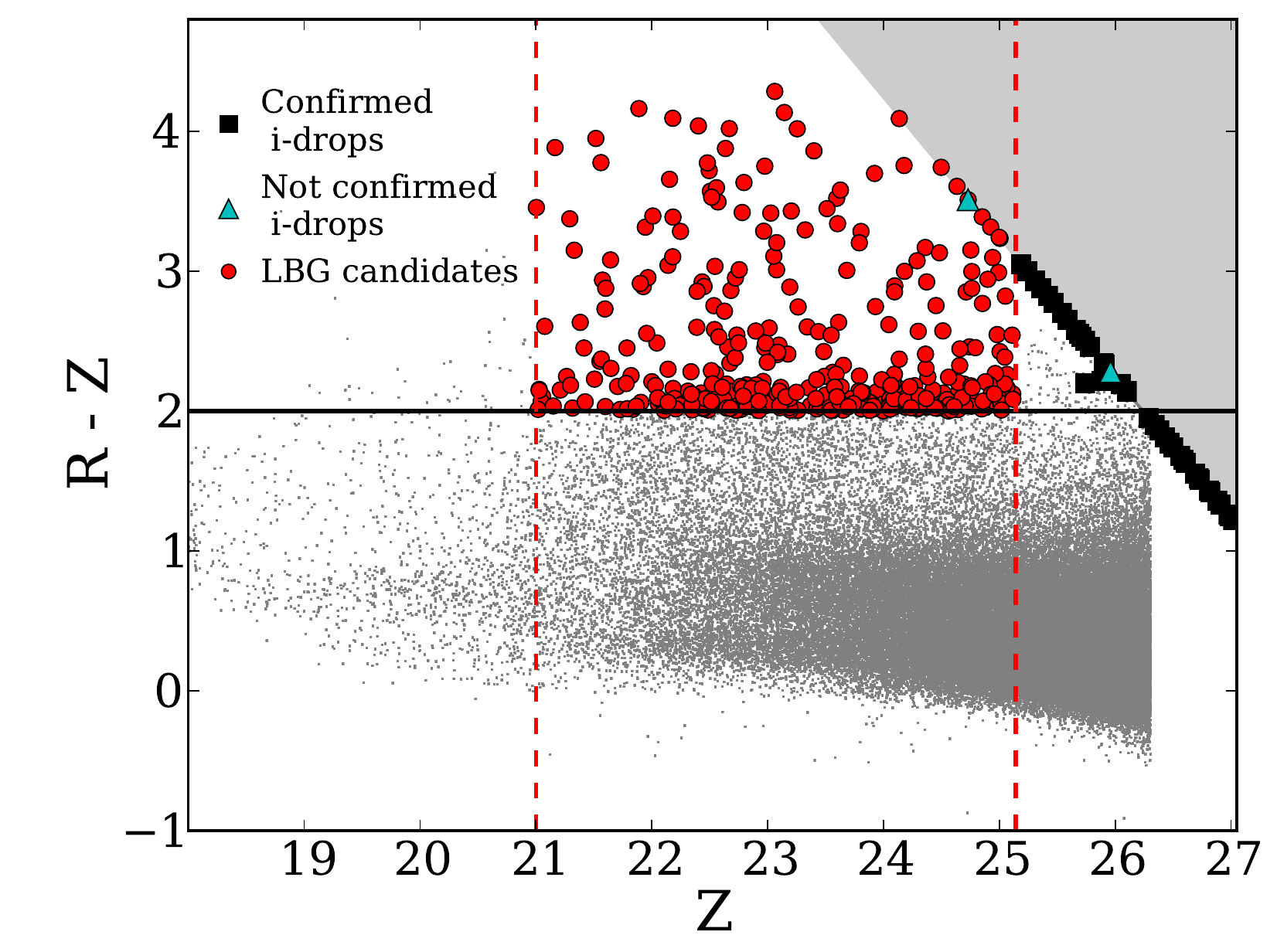}
\caption{Color--magnitude diagram for the Subaru Deep field (SDF). Gray points represent all the objects in the field.
The black squares correspond to the 42 spectroscopically confirmed $i$-dropout galaxies 
in \cite{nag04,nag05,nag07}, \cite{ota08}, \cite{jia11}, \cite{tos12}, and \cite{jia13}.
The cyan triangles are $i$-dropout candidates reported by \cite{nag07} that could not be confirmed or rejected as 
galaxies at $6.0 < z < 6.5$ by their spectra. When the $i$-dropout galaxies are not in our SDF catalog, 
the $Z$-band magnitude of the discovery papers are used.
The horizontal solid line is our criteria for selecting LBGs with a break larger than $R-Z=2$.  The vertical 
dashed lines are the brighter and fainter magnitudes considered for our candidates, $Z = 21$ and $Z=25.14$, respectively (see the text).
The gray region shows $R$-band magnitudes fainter than $2\sigma$ limiting magnitude ($R=28.24$).  The 370 LBG candidates
in this field are indicated with red circles.\label{fig:cmd_bf}}
\end{figure}

Figures \ref{fig:cmd_qf} and \ref{fig:cmd_bf}
present the color--magnitude diagram and the selection criteria of the $Z$-band
selected objects in the J0203 and SDF fields respectively. In red 
circles we show the LBG candidates detected in these fields.
There were 20 LBG candidates in the quasar field, and  
this sample contained the 8 NB-selected LBGs from Section \ref{ss:LBGsel}. We 
found 370 LBG candidates in the SDF.
Scaling the LBG candidates in the SDF to our effective area, the expected number of
candidates is $\sim 19$. Additionally, we placed $100{,}000$ FORS2 fields-of-view at random
positions in the SDF. Counting how many LBG candidates fell in each FORS2 field, resulted in 
a Gaussian distribution with mean $\mu=16$ and standard deviation $\sigma = 5$.
We concluded that the density of LBG candidates in the quasar field is consistent with that 
of a blank field.

Figure \ref{fig:cmd_bf} also presents the compilation of known $i$-dropout galaxies in the SDF.
With our depth we could not detect any of the 42 spectroscopically confirmed
galaxies. We were only able to recover the brightest candidate of \cite{nag07}, a source that
could not be confirmed or ruled out as a galaxy at $6.0<z<6.5$ by its spectrum. Additionally,
if we used the extra information of the SDF provided by the $i$ band, we noticed that the majority
of the objects that we selected as possible LBGs seemed to be contaminants with a rather
shallow slope in the $R-i$, $i-Z$ colors instead of having a sharp break.

From this analysis, we concluded that our LBG sample is very likely to be 
highly contaminated. We therefore cannot reliably estimate the (over)density of LBGs around
our quasar. In the following sections we thus focus primarily on the LAE results.

\subsection{Photometric Properties of the LAEs} \label{sec:LAEprop}

We assume a simple model where the LAE spectrum consists of the Ly$\alpha$ line
with flux $F_{{\rm Ly}\alpha}$, a continuum flux density redward of Ly$\alpha$ with strength $C$
and a power law slope $\beta$ ($f_\lambda \propto \lambda^{\beta}$). Then, the flux density in the 
NB ($f_{\lambda,{\rm NB}}$) and $Z$ ($f_{\lambda,Z}$) can be written as:
\begin{eqnarray}
 f_{\lambda,{\rm NB}} &=& C \lambda_{{\rm eff, NB}}^\beta + F_{{\rm Ly\alpha}} / \Delta \lambda_{{\rm NB}} \\
 f_{\lambda,Z} &=& C \lambda_{{\rm eff}, Z}^\beta\mbox{,}
\end{eqnarray}

\noindent with $\lambda_{{\rm eff}, i}$ the effective wavelength of the $i$ filter and
$\Delta \lambda$ the width of the filter. \cite{dun12a} found that high redshift galaxies
($5<z<7$) have an average slope of $\beta \simeq 2$ and this does not show a significant trend with either 
redshift or $M_{\rm UV}$. Assuming a flat continuum, i.e., $\beta = -2$, we solve the previous $2 \times 2$ linear system and
estimate the line flux and continuum flux density.

To calculate $L_{\rm Ly\alpha}$ and equivalent width, we use the relations:
\begin{eqnarray}
L_{\rm Ly\alpha} &=& 4\pi d_L^2(z=5.7) F_{\rm Ly\alpha} \\
{\rm EW_{obs}} &=& \frac{F_{\rm Ly\alpha}}{C(\lambda_{\rm Ly\alpha}(1 + z)^\beta)}
\end{eqnarray}
\noindent with $d_L$ the luminosity distance and $\lambda_{\rm Ly\alpha}$ the wavelength of the Ly$\alpha$ line.
The rest-frame equivalent width is given by $\rm EW_0 = EW_{obs} / (1 + z)$.

To calculate star formation rates (SFRs), we follow the approach of \cite{ouc08}:
\begin{equation}
 \mbox{SFR}(M_\sun \mbox{ yr}^{-1}) = \frac{L_{\rm Ly\alpha}}{1.1 \times 10^{42} \,\mathrm{erg}\, \mathrm{s}^{-1}}
\end{equation}
\noindent where they used the relation of H$\alpha$ luminosity and star formation rate \citep{ken98}
and assumed the standard case B recombination factor of 8.7 for the Ly$\alpha/$H$\alpha$ luminosity ratio \citep{bro71}.

We calculate the input parameters for the filters used in this project, 
taking into account both the filter curves and the quantum efficiency of the CCD:
$\lambda_{\rm eff, NB} = 8150.2$ \AA, $\lambda_{{\rm eff}, Z} = 9180.2$ \AA, 
and $\Delta \lambda_{\rm NB} = 119$ \AA. 
The star formation rates and rest-frame equivalent widths for our two LAE candidates from
Section \ref{ss:laesel} are
\mbox{$(\mbox{SFR},\, {\rm EW_0}) = (6.0 \pm 1.3 \; M_\odot \, \mathrm{yr}^{-1},\, 47^{+222}_{-16} \; \mbox{\AA})$}
and 
\mbox{$(\mbox{SFR},\,{\rm EW_0} ) = ( 1.7 \pm 0.9 \; M_\odot \, \mbox{yr}^{-1}, \,  \geq  14 \; \mbox{\AA})$}
for LAE 1 and LAE 2, respectively (see Figures \ref{fig:field} and \ref{fig:stamps}).

The properties derived for LAE 1 are within the expected range for typical LAEs 
\citep[see][ SFR = $6.2^{+2.7}_{-1.9} \; M_\odot \, \mathrm{yr}^{-1} $ for $L^*$ LAEs
at $z=5.7$]{ouc08}, whereas for  LAE 2,  the SFR  is lower than for typical LAEs and 
the equivalent width is only a lower limit.

\subsection{Black Hole Mass of the Quasar}

High-redshift quasars hosting black holes with masses $> 10^9 M_\odot$ are thought to 
be hosted by supermassive dark matter halos as suggested by their extremely
low comoving density and large black hole masses \citep[e.g.,][]{vol06}. No black 
hole measurement exists for J0203. Typically, black hole masses of high-redshift quasars are 
estimated using the single-epoch black hole mass estimators based on continuum luminosities
and broad emission lines such as Mg$\,${\sc ii} and C$\,${\sc iv}  
\citep[e.g.,][]{jia07, kur07, der11}.
For a source at $z=5.72$, the Mg$\,${\sc ii} line appears at 1.88 $\mu$m, 
in the middle of the strong telluric absorption band at $\sim 1.9\, \mu$m.
Moreover, since J0203 is a broad absorption line quasar,
the C$\,${\sc iv} line is strongly absorbed (see Figure 7 in Mortlock et al. 2009). That is why
we cannot use these lines to measure the black hole mass of J0203.
Instead, we estimate the quasar Eddington luminosity in order to determine the
black hole mass, assuming that the central black hole accretes at the 
Eddington limit. We measure the 
flux at $\lambda = 1350\, $\AA, yielding a monochromatic luminosity of 
$3.7 \times 10^{46}$ erg s$^{-1}$. Using the bolometric correction of 3.81 from 
\cite{ric06}, we obtain a bolometric luminosity of $1.4 \times 10^{47}$ erg s$^{-1}$
with 0.2 dex of uncertainty. \cite{der11} found that $L_{\rm bol} / L_{\rm Edd} \sim 0.43$ with 
a scatter of 0.2 dex for luminous ($L_{\rm bol} \gtrsim 10^{47}$ erg s$^{-1}$) $z>4$ quasars.
Assuming this ratio applies to J0203, our estimate of the black hole mass is 
$2.5 \times 10^{9}$ $M_\odot$ with 0.3 dex uncertainty. This mass is comparable
to the typical black hole masses of the brightest SDSS quasars, thus it is plausible 
that J0203 resides in a massive dark matter halo.

\section{Discussion and Conclusions} \label{sec:discussion}

We present a deep study of the LAE and $R-Z$ dropout population centered on 
the $z=5.72$ quasar J0203 with an estimated black hole mass 
of $\sim 2.5 \times 10^{9}$ $M_\odot$.
The redshift of the quasar enables a LAE narrow-band search in its vicinity.
We detect 2 LAE  and 20 $R-Z$ dropout candidates (the number of dropout candidates
decreases to 8 if we use the narrow band for detection and selection).
The LAE sample spans a very narrow redshift range around the quasar: $5.66<z<5.75$.
On the other hand, the $R-Z$ dropout sample spans a larger redshift range as is typical for
dropout selections: $5.2 \lesssim z \lesssim 6.8$ ($5.2 \lesssim z \lesssim 5.8$, if the 
narrow-band filter is added to the selection).

Comparing our LAE counts with the luminosity functions of two LAE surveys that do not
target a quasar \citep{ouc08, hu10}, we find
that our number counts are consistent with what is observed in blank fields.

The number of $R-Z$ dropout candidates in the field of J0203 is consistent with the 
expected number from the SDF blank field. However, we find that with our depth and using only
two broad bands for the selection, it is very likely that our candidates are highly contaminated.
This implies that we cannot establish a reliable estimate of the (over)density of LBGs around
J0203.

There are various ways to explain the non-detection of an overdensity of LAEs around the quasar.

\begin{enumerate}
 \item Low number statistics may prevent us from detecting a possible overdensity of emission
 line galaxies.  If there were an overdensity it  would be at least a factor of $\sim 3$--$5$,  
 based on environmental studies of radio galaxies. Since we are studying only  one quasar,
 it could be that we are targeting a rare source and  that other quasars still harbor  
 overdensities in their surroundings.  This is certainly a possibility, but we believe that 
 it is unlikely given the relatively low  counts of LAEs and LBGs toward other $z \sim 6$ 
 quasars studied to date.
 
 \item The lack of neighbors in the immediate vicinity could be due to mergers of galaxies
 in the halo that hosts the quasar. However, we believe that such mergers should not prevent us 
 from detecting an overdensity based on the protoclusters found around radio galaxies 
 \citep[e.g.,][]{ven07a} and the recently spectroscopically confirmed LBGs in the vicinities
 of quasars at $z\sim 5$ found by \cite{hus13}, which in principle should have been
 affected in the same manner.

 \item The strong ionizing radiation from the quasar may prevent galaxy formation in its 
 surroundings. There are several studies that support this claim \citep[e.g.,][]{bru12, fra04}
 and therefore it would be important to understand and quantify 
 this effect in order to consider it in simulations. Indeed, our results present evidence to
 support this idea because none of our candidates are at a distance closer than $\sim 4.5$ 
 comoving Mpc ($\sim 0.7$ physical Mpc), as can be seen  in Figure \ref{fig:field} and 
 Table \ref{tbl-1}. This is similar to the results presented by \cite{kas07} and \cite{uts10}, 
 i.e., that galaxies avoid the vicinity of a quasar within distances
 of $\sim 4.5$ comoving Mpc and $\sim 15$ comoving Mpc, respectively. 

Yet, the quasar ionization hypothesis  fails to explain 
why apparent overdensities are found in the vicinities of other high-redshift quasars
\citep[e.g.,][]{sti05}. Further, we believe that is hard to reconcile a hypothesis 
in which the radiation affects the environment isotropically, with the unified AGN model
where quasars have two collimated beams.
 However, contrary to the ionization hypothesis, \cite{can12} suggest that
the radiation of the quasar may enhance the number of LAEs in its surroundings.
They show that gas-rich objects with little or no associated star formation, known as
proto-galactic clouds or dark galaxies, can be detected thanks to the Ly$\alpha$ 
fluorescence induced by a nearby quasar. One of the main characteristics of these
sources is a high Ly$\alpha$ equivalent width \mbox{($\rm EW_0 > 240$ \AA).}
 A detailed comparison with \cite{can12} is challenging because of the different
redshifts and because the Lyman limit luminosity of our quasar, calculated following 
\cite{hen06}, is about one order of magnitude smaller than the luminosity of the 
hyperluminous quasar (HLQSO) observed by \cite{can12}. Additionally, since J0203 is 
a broad absorption line quasar \citep{mor09}, it is possible that the amount of escaping ionizing radiation
is lower than expected in typical quasars.

Nevertheless, in our study, we do not find such fluorescent clouds. It should be noted that in our work, 
LAE 1 has an $\rm EW_0 =  47^{+222}_{-16}$ \AA $\,$ which has a 4\% probability of having 
\mbox{$\rm EW_0 > 240$ \AA} $\,$based on the uncertainties of the measured
fluxes and LAE 2 only has a lower limit of \mbox{$\rm EW_0 \geq 14$ \AA.}
    
 \item High-redshift quasars may not reside in the most massive halos. It is expected that the relation between black hole mass and
 halo mass that arises from combining the local relations between black hole and bulge mass on the one hand, and the relation between galaxy
 and halo mass on the other, is significantly different (both in slope and in scatter) at high redshift. Evidence for this has been found 
 recently at $z \sim 2.7$ by  \cite{tra12} who studied fifteen HLQSOs ($\gtrsim10^{14} L_\odot$; $M_{1450 }\simeq -30$) that
 are associated with galaxy overdensities. The authors conclude that HLQSOs do not require environments very different from their much less 
 luminous quasar counterparts and are not being hosted by rarer dark matter halos. However, if this applies to the $z \sim 6$ quasars as well, 
 one would have to explain their extremely low space density. One possibility is that, due to selection effects, the bright quasars 
 thus far discovered are biased toward more average-sized halos that are inhabited by very massive black holes.

 \item High-redshift quasars may not always reside in highly overdense, large-scale environments. All observational evidence considered, 
 there appears to be a trend for $z \sim 6$ quasars to sit in low-to-average galaxy environments on megaparsec scales. $\Lambda$CDM predictions 
 have shown that the most massive halos at high redshift do not invariably grow into the most massive ones at low redshift. Moreover,
 the most massive present-day structures (clusters) originated primarily from regions that were  overdense on very large scales
 (tens of megaparsec) at $z \sim 6$ \citep{del07, ove09, ang12}. Although few studies 
 have probed out to such large radii and with 
 significant sensitivity, the lack of companion galaxies on megaparsec scales may indicate that the quasars do not pinpoint progenitors of 
 the most massive clusters even if the quasars themselves are hosted by very massive halos. 
It is also possible that quasars reside in large-scale overdense environments,
 but just not in the center of them. Studies with larger fields-of-view are required to test
 this hypothesis.
   
\end{enumerate}

One question that emerges naturally is why radio-galaxy-based protocluster surveys have been much more successful than any quasar-based protocluster
survey.  From a theoretical perspective, N. Fanidakis \& Orsi (2013, in preparation) suggest that radio galaxies and quasars are observational 
manifestations of different accretion states, which are determined by the accretion and black hole properties and the large scale environment. 
Quasars accrete vast amounts of gas and form in the gas-rich environments of intermediate mass ($\sim10^{12}\, M_{\odot}$) dark-matter 
halos \citep{fani13}. Radio galaxies, in comparison, are powered by rapidly spinning black holes that typically inhabit 
the centers of the most massive ($\gtrsim10^{13}\,M_{\odot}$) dark matter halos in the 
universe and accrete low-density gas \citep[for the properties of their AGN model; see][]{fani11,fani12}. The largest concentrations of baryons are expected 
to be found within these halos and thus, radio galaxies can be used to pinpoint the location of large galaxy overdensities. 
The suggestion of a discrete halo environment of quasars and radio galaxies is also supported by clustering analysis of AGNs in the low-redshift
universe \citep[$0<z<2$; e.g.,][]{ros09,don10,wak11}.

More detailed studies of the environment and of the host galaxy of high-redshift quasars are needed to clarify the picture of overdensities
around quasars. Future observational campaigns targeting high-redshift quasars either at  $z=5.7$, $z=6.6$, or $z=7.0$
(corresponding to the few gaps of OH night-sky emission bands where narrow-band filters can be
most effectively targeted to detect Ly$\alpha$ emission) will be crucial to increase the sample
and to quantify possible object-to-object variations.

\acknowledgments
Based on observations made with ESO Telescopes at the La Silla Paranal Observatory
under program ID 385.A-0030(A).
E.B. thanks N. Fanidakis, J. Hennawi, and M. Tanaka for useful discussions about this work,
 M. Maseda for proof reading a preliminary version of this manuscript, and
the IMPRS for Astronomy \& Cosmic
Physics at the University of Heidelberg. The plots in this publication 
were produced using Matplotlib \citep[][\url{http://www.matplotlib.org}]{hun07}.

{\it Facility:} \facility{VLT (FORS2)}.

\end{document}